\documentclass[prd,onecolumn,preprintnumbers,nofootinbib]{revtex4}
\usepackage[a4paper, hdivide={1.91cm,,1.165cm}, vdivide={1.83cm,,2.6cm}]{geometry}

\usepackage{amstext,amssymb}
\usepackage{amsmath}
\usepackage{graphicx}
\usepackage[hyperfootnotes=false]{hyperref}
\usepackage{xspace}
\usepackage{color}
\usepackage{units}
\usepackage{slashed} % feynman slash notation via \slashed{p}

\begin{document}
%%%%%%%%%%%%%%%%%%%%%%%%%%%%%%%%%

\title{Low scale trinification symmetry through effects of dimension-5
operators in $E_6$ grand unified theory with and without supersymmetry}

%\preprint{ULB-TH/15-10}

\author{Chandini  \surname{Dash}}
\email{dash25chandini@gmail.com}
\affiliation{Department of Physics, Berhampur University, Odisha-760007, India}

\author{Snigdha  \surname{Mishra}}
\email{mishrasnigdha60@gmail.com}
\affiliation{Department of Physics, Berhampur University, Odisha-760007, India}

%%%%%%%%%%%%%%%%%%%%%%%%%%%%%%%%%%
%%%%%%%%%%%%%%%%%%%%%%%%%%%%%%%%%%
\begin{abstract}
We examine $E_6$ GUT models (both SUSY and non-SUSY) with D-parity violating intermediate trinification symmetry $\left(SU(3)_C\otimes SU(3)_L\otimes SU(3)_R\right)(g_{3L}\neq g_{3R})$ where the D-parity breaking is achieved by quantum gravity effect through dimension-5 operators. Inclusion of the gravitational effect significantly lowers the intermediate scale, being as low as $10$ TeV. However with a given choice of the intermediate mass $M_I$, the gauge unification mass $M_U$ as well as the GUT coupling constant are unaffected by the effects. The predicted value of $M_U$ is shown to be compatible with the accessible limit of the proton lifetime in presence of additional particles i.e. color octet scalars like $(8,2,\frac{1}{2})$ or $(8,1,0)$ and electroweak triplet fermion $(1,3,0)$ at the lower scales. The presence of color octet scalar with mass of the order of TeV may suppress the production of the Higgs boson through gluon fusion. The $SU(2)_L$ triplet fermion may behave like a stable dark matter. With reference to cosmological issue, it is expected that no strings or walls bounded by strings associated with D are generated, since the discrete D-parity is broken at the GUT scale.

\vspace*{-1.0cm}
\end{abstract}

% \pacs{12.60.Cn}
%12.60.Cn 	Extensions of electroweak gauge sector
%95.35.+d 	Dark matter

%%%%%%%%%%%%%%%%%%%%%%%%%%%%%%%%%
%%%%%%%%%%%%%%%%%%%%%%%%%%%%%%%%%
\maketitle

%%%%%%%%%%%
\section{Introduction}
\label{sec:intro}

The Standard Model(SM), the most elegant theory of particle physics, needs to be extended in a well motivated way to explain some of the experimental predictions of LHC. The trinification model based on the gauge group $SU(3)_C\otimes SU(3)_L\otimes SU(3)_R$ \cite{Babu:1985gi,Nishimura:1988fp,Sayre:2006ma,Cauet:2010ng,Stech:2014tla,Hetzel:2015cca,Babu:2017xlu} is an interesting extension of the Standard Model to comply with its unsolved issues like detection of dark matter candidate, ultralight neutrino and to explain the matter-antimatter asymmetry. Being embedded maximally in the $E_6$\cite{Gursey:1975ki,Shafi:1978gg} grand unified theory, the model automatically contains all the nice features of the conventional GUT models like $SU(5)$ and $SO(10)$ etc. The motivation for trinification lies in the unified description of both strong and electroweak interactions, while incorporating nice features of the conventional left-right symmetric (LRSM) models. Many $E_6$ GUT models with D-parity (D$_{LR}$ being a discrete left-right symmetry) \cite{Chang:1983fu,Chang:1984uy} conserving trinification symmetry have been discussed \cite{Stech:2003sb,Wang:2011eh,Chakrabortty:2017mgi,Dash:2019bdh,Dash:2020jlc} with phenomenological prediction.

However the present study is an attempt to revisit the models with broken D-parity. The D-parity breaking $(g_{3L}\neq g_{3R})$ can be achieved by quantum gravity effect through non-renormalizable dimension-5 operators \cite{Shafi:1983gz,Hill:1983xh,Rizzo:1984mk,Patra:1991dy,Chakrabortty:2008zk,Huang:2017uli,Dash:2019bdh}. In fact the presence of these operators show the remnants of Planck scale physics, which in turn imply the unification of the GUT model with gravity. Thus the operators initiate gravitational correction to the gauge couplings and hence affect the unification mass scale and GUT coupling constant. Keeping this in view, the prime objective of the present work is to search for viable  $E_6$ GUT models with intermediate trinification symmetry, in presence of gravitational effect. Since supersymmetry(SUSY) GUTs have been at the centre of attention for a number of attractive features, it is tempting to explore the above mentioned possibility, with and without SUSY. In the present work, we show that inclusion of gravitational corrections can lower the intermediate scale, bringing it even to the range of a few TeV.  

It has been established through a theorem \cite{Dash:2019bdh,Dash:2020jlc} that in all grand unified theories(GUTs) having trinification symmetry ($G_{333D}$) as the highest intermediate symmetry, the electroweak mixing angle $\sin^2\theta_{W}$ and the intermediate mass scale $M_I$ have vanishing contributions due to one (two) loop and gravitational as well as threshold corrections. It is noteworthy to mention that the D-parity conservation is solely responsible for this vanishing correction. However, with broken D-parity, the scenario changes i.e. with the impact of gravitational correction, $M_I$ changes. The gauge unification mass $M_{U}$ as well as the GUT coupling constant $\alpha_{G}$ are unaffected by gravitational effects, with the intermediate mass $M_{I}$ as a free parameter. An important observation of the present work shows that, the gravitational correction, as a bonus permits low intermediate scale with observable effects. It is known that, GUT model with high intermediate scale has the disadvantage that the neutrino seesaw related new physics is well hidden form low energy and collider probes thereby making it untestable. Thus our main focus is to achieve low intermediate scale through non-renormalizable operators. Here the predicted value of $M_U$ is shown to satisfy twin requirements that it is in agreement with the proton decay bound and is within the upper limit set by the Planck mass as well as all couplings remain perturbative.

The paper is organised as follows. In the next section we give a brief analysis for the requirements of perturbativity up to Planck scale. We discuss the general framework for the models along with the analytical expression of the unification mass, inverse GUT coupling constant in section-III. In section-IV, we give the numerical estimations for both SUSY and non-SUSY models. The last section is devoted to a discussion  on the phenomenological implications of the numerical results.

 \section{Perturbative Criteria for an $E_6$ GUT model}
 
The Higgs sector of a GUT model plays a very crucial role in accomplishing two important tasks like breaking the GUT symmetry down to the Standard Model and to give mass to the matter sector in tune with the current experimental data. In general one can have many choices of Higgs, but the minimal choice is always better, provided the obtained unification mass is consistent with proton decay constraint and the GUT coupling constant remains perturbative till Planck scale. In the present context of $E_6$ GUT model, which is rather a larger group with many exotic particles, unlike the conventional $SO(10)$ and $SU(5)$ GUTs, the perturbative requirement \cite{Chang:2004pb,Kopp:2009xt} needs to be examined for a successful modeling. It is obvious that, for sufficiently large particle content, the non-perturbative regime is reached at relatively low scales. Specifically, addition of new (non-singlet) matter/scalar particles in a given model for the sake of unification, always affects the perturbative condition. It is particularly problematic in non-minimal SUSY models. For the sake of completeness, in the present context of SUSY and non-SUSY $E_{6}$ GUTs, we analyse this by locating the the Landau pole$(\mu_0)$, where the inverse GUT coupling constant $\alpha_G^{-1}$  vanishes. Here, we limit to one-loop R.G. equation to find out the pole, for a given set of gauge, matter and Higgs particles belonging to $E_6$. To derive this constraint, we note the (one-loop) RG equation for the unified gauge coupling constant within the mass range $M_U$ and $\mu_0$ where $M_U$ is the unification mass scale and $(\mu_0)$ is the position of the Landau pole.
 
\begin{eqnarray}
\alpha_G^{-1}(\mu_0)= \alpha_G^{-1}(M_U)-\frac{{b}}{2 \pi} {\large \ln}\left(\frac{\mu_0}{M_U}\right)
\label{eqn:RG}
\end{eqnarray}

By using the standard formula, the one-loop beta coefficients $b$ is determined by the particle content of the model with contribution from  $b_{gauge}$, $b_{Higgs}$ and $b_{matter}$ of the GUT group $E_6$. Putting $\alpha_{G}^{-1}=0$, at the Landau Pole $\mu_0$, we have,  

\begin{eqnarray}
\mu_0= M_{U} \exp \Bigg[\frac{2\pi}{b} \alpha_G^{-1}(M_U) \Bigg]
\label{eqn:landau}
\end{eqnarray}

\begin{table}[h!]
% \vspace*{-0.6119cm}
\begin{tabular}{||c|c||}
\hline
Representations & beta coefficients $(b)$ \\
\hline
$27$ & $3$ \\
$78$ & $12$ \\
$351$ & $75$ \\
$351^{'}$ & $84$ \\
$650$ & $150$ \\
\hline
\end{tabular}
\caption{Beta coefficient of $E_6$ representations for the GUT gauge coupling evolution}.
\label{irred}
\end{table}

Using the values of beta coefficients from Table-\ref{irred}, we can locate the pole position corresponding to various choices of Higgs sectors. For $E_6$ GUTs with three generations of matter fields, we note that, $b_{gauge}=(-44)$ and $(-36)$ and $b_{matter}=6$ and $9$ for non-SUSY and SUSY cases respectively. 
Thus we have,

\begin{eqnarray}
  b=&&-38+\left(\frac{1}{3} \right)b_{Higgs}; \hspace*{+3.0cm} non-SUSY\nonumber\\
  b=&&-27+b_{Higgs}; \hspace*{+4.0cm} SUSY
  \label{eq:b}
\end{eqnarray}

In the present context of $E_6$ GUTs (with one intermediate symmetry) for  different possible choices of the Higgs sector, the Landaue pole position$(\mu_0)$ can be estimated. For a rough estimate one can have a limit for choosing the appropriate Higgs in the GUT models. Putting $\mu_0=10^{18.38}$ GeV (the reduced Planck scale), $M_{U}=10^{16}-10^{17}$ GeV in equation (\ref{eqn:landau}), the perturbativity till $\mu_0$ demands, 
\begin{eqnarray}
 \frac{b}{\alpha_{G}^{-1}(M_{U})}=1.146-1.977
\end{eqnarray}
 
 where $b$ is given by equation (\ref{eq:b}). Usually, in SUSY $E_6$ models, due to large particle contents with high beta coefficients, it is difficult to push the Landaue pole closer to the Planck scale. However in presence of gravitational correction, it may be possible to obtain high values of the unification scale and perturbative gauge coupling up to the Planck scale by appropriate choice of the correction parameter. In the subsequent section we focus on these possibilities.

\section{The Model Framework}
% \subsection{The Model}
\label{sec:model-I}
We consider here an $E_6$ GUT (with and without SUSY) with D-parity violating intermediate trinification symmetry $SU(3)_C \otimes SU(3)_L \otimes SU(3)_R$ where D-parity is spontaneously broken at the unification mass scale $M_U$. The corresponding breaking pattern is given as,

\begin{eqnarray}
&& E_6 (\otimes SUSY)\stackrel{M_U}{\longrightarrow}(\mathbb{G}_{333}) SU(3)_C\otimes SU(3)_L\otimes SU(3)_R (g_{3L}\neq g_{3R}) (\otimes SUSY)\nonumber \\ 
	&&\stackrel{M_I}{\longrightarrow}(\mathbb{G}_{321})SU(3)_C\otimes SU(2)_L\otimes U(1)_{Y} (\otimes SUSY)\nonumber\\
	&&\stackrel{M_Z}{\longrightarrow}(\mathbb{G}_{31}) SU(3)_C\otimes U(1)_{Q} 
	\label{eq:Model}
	\end{eqnarray}

The first step of spontaneous symmetry breaking from $E_6$ GUT to $G_{333}$--is achieved by giving a GUT scale VEV to D-parity odd singlet scalar $\phi (1,1,1)$ contained in $650_H\subset E_6$ leading to $g_{3L}\neq g_{3R}$.
The next stage of symmetry breaking i.e. from $\mathbb{G}_{333} \to \mathbb{G}_{\rm 321}$ is done by assigning a non-zero VEV to the $\mathbb{G}_{\rm 321}$ neutral component of either $27_H$ or $351^{'}_H$ . 
The last stage of symmetry breaking i.e. SM to low energy theory $(G_{31})$ is done by assigning a non-zero VEV to SM Higgs doublet contained in $27_H \subset E_6$. 
\\
 At the unification mass scale we introduce non-renormalizable  
dimension-5 operator 
\begin{eqnarray}  
&&\mathbb{L}_{NRO}=-\frac{\eta}{4 M_{G}} {Tr} \Big(F_{\mu\nu} {\Phi_{650}} F^{\mu\nu}\Big)
\label{eq:NRO}
\end{eqnarray}

which induces gravitational corrections. This $\mathbb{L}_{NRO}$ may arise due to spontaneous compactification effects of extra dimensions or due to the quantum gravity. Here in equation (\ref{eq:NRO}), the scale $M_{G}\simeq M_{Pl} \simeq 10^{19}$ GeV, if it is due to quantum gravity effect. However, if it emerges as a result of compactification of extra dimensions then $M_{G} \leq M_{Pl}$. $ \eta $ is a dimensionless parameter and $F_{\mu\nu}$ is the field strength tensor. As has been mentioned before, $ \Phi_{650} $ is the odd singlet scalar which takes the VeV at the mass scale $M_U$.

\vspace{-10pt}
\begin{eqnarray}
 & &\hspace*{-1.0cm} \langle {\Phi_{650}} \rangle=\frac{\langle\phi^0\rangle}{\sqrt6} {diag} \left\{\underbrace {\epsilon_{3C}}_{9},\underbrace{\epsilon_{3L}}_{9},\underbrace{\epsilon_{3R}}_{9}\right\} \nonumber
 \label{eq:650VEV}
\end{eqnarray}
with ${\epsilon} = \frac{1}{\sqrt{6}}\frac{\eta\, \langle \phi^0 \rangle}{M_{G}}$, we have ${\epsilon_{3C}} = 0,
    {\epsilon_{3L}} ={\epsilon}=-{\epsilon_{3R}}$, which breaks D-parity.
 This assigned VeV modify the boundary condition of the gauge couplings $\alpha_{3C}$, $\alpha_{3L}$ and $\alpha_{3R}$ given by,
\begin{eqnarray}
 &&\alpha_{3C} (M_U)=(1+\epsilon)\alpha_{3L} (M_U)=(1-\epsilon)\alpha_{3R} (M_U)=\alpha_{G}(M_U)
   \end{eqnarray}

where $\alpha_{G}$ is the GUT coupling constant.
Now using this modified boundary conditions for the gauge couplings, we can express the RG equation at different mass ranges $M_{Z}-M_{I}$ and $M_{I}-M_{U}$.
 As has been mentioned before, within the mass scale $M_I$ to $M_U$, the $L_{NRO}$ will be operative in the RGE. We may note here that in order to achieve unification in $E_6$ GUT models with one intermediate symmetry with the check point of proton decay constraint, additional light particles may be required. In the present frame work, corresponding to different models, additional light scalars/fermions (with masses $M_1$ and $M_2$) are introduced at lower scales within the mass $M_I$ to $M_Z$. These additional particles may be correlated with consistent phenomenology that may be accessible in future collider, which will be discussed in the later section. With the above input the evolutions of the Standard Model gauge couplings can be written as

\begin{eqnarray}
  &&\alpha^{-1}_{3C} (M_Z)=\alpha^{-1}_{G} + \frac{b_{3C}}{2 \pi} {\large \ln}\left(\frac{M_1}{M_Z}\right)+ \frac{b^{1}_{3C}}{2 \pi} {\large \ln} \left(\frac{M_2}{M_1}\right) + \frac{b^{2}_{3C}}{2 \pi} {\large \ln} \left(\frac{M_I}{M_2}\right)
 + \frac{b^{U}_{3C}}{2 \pi} {\large \ln} \left(\frac{M_U}{M_I}\right) 
 \label{eqn:3C-I}
    \end{eqnarray}
 \begin{eqnarray}
&&\alpha^{-1}_{2L} (M_Z)=\alpha^{-1}_{G}(1+\epsilon) + \frac{b_{2L}}{2 \pi} {\large \ln}\left(\frac{M_1}{M_Z}\right)+ \frac{b^{1}_{2L}}{2 \pi} {\large \ln} \left(\frac{M_2}{M_1}\right) + \frac{b^{2}_{2L}}{2 \pi} {\large \ln} \left(\frac{M_I}{M_2}\right)
 + \frac{b^{U}_{3L}}{2 \pi} {\large \ln} \left(\frac{M_U}{M_I}\right)
 \label{eqn:2L-I}     
 \end{eqnarray}
  \begin{eqnarray}
 &&\alpha^{-1}_{Y} (M_Z)=\left(1-\frac{3\epsilon}{5}\right)\alpha^{-1}_{G}  + \frac{b_Y}{2 \pi} {\large \ln}\left(\frac{M_1}{M_Z}\right)+ \frac{b^{1}_{Y}}{2 \pi} {\large \ln} \left(\frac{M_2}{M_1}\right)+ \frac{b^{2}_{Y}}{2 \pi} {\large \ln} \left(\frac{M_I}{M_2}\right)
 \nonumber \\
 &&\hspace*{3cm}+\left( \frac{\frac{1}{5} b^{U}_{3L} 
  + \frac{4}{5} b^{U}_{3R}}{2 \pi}\right) {\large \ln} \left(\frac{M_U}{M_I}\right) 
  \label{eqn:Y-I}
 \end{eqnarray}
 
 where $b_{i}$ and $b_{i}^{1,2}$ with $i=3C, 2L, 1Y$ are the one-loop beta coefficients between the mass range $M_Z$ to $M_1$, $M_1$ to $M_2$ and $M_2$ to $M_I$ respectively. Similarly $b_{i}^{U}$ with $i=3C, 3L, 3R$ are the one-loop beta coefficients between the mass range $M_I$ to $M_U$.

We now follow the standard procedure to obtain the analytical expression for the unification mass scale $M_U$, the inverse GUT coupling constant $\alpha_G^{-1}$ and electroweak angle $\sin^2\theta_W$, given as,

\begin{eqnarray}
 &&{\large \ln}\left(\frac{M_U}{M_Z}\right) = 
 \frac{1}{4\Big({b_{3L}^{U}}+{b_{3R}^{U}}-2{b_{3C}^{U}} \Big)} \Bigg[16\pi \left(\frac{3}{8}\alpha^{-1}_{em}(M_Z)-\alpha^{-1}_{s}(M_Z)\right)\nonumber \\
 &&\hspace*{+1.40cm}-\sum_{\alpha=1,2}\Bigg\{\Big(8\Delta_{3C}^{\alpha}-3\Delta_{2L}^{\alpha}-5\Delta_{Y}^{\alpha}\Big) {\large \ln}\left(\frac{M_{\alpha}}{M_Z}\right)\Bigg\}\nonumber \\
 &&\hspace*{+1.40cm}
+\Bigg\{4\Big({b_{3L}^{U}}+{b_{3R}^{U}}-2{b_{3C}^{U}} \Big)+\Big(8{b_{3C}^{2}}-3{b_{2L}^{2}}-5{b_{Y}^{2}} \Big)
 \Bigg\}{\large \ln}\left(\frac{M_I}{M_Z}\right)\Bigg]
 \label{rel:MU}
\end{eqnarray}

\begin{eqnarray}
&&\alpha_{G}^{-1} =\frac{1}{4\Big({b_{3L}^{U}}+{b_{3R}^{U}-2{b_{3C}^{U}}} \Big)}
 \Bigg[4\Big({b_{3L}^{U}}+{b_{3R}^{U}}\Big)\alpha^{-1}_{s}(M_Z)-3{b_{3C}^{U}}\alpha^{-1}_{em}(M_Z)
 \nonumber \\
 &&\hspace*{+1.00cm}
 +\frac{1}{2\pi}\sum_{\alpha={1,2}}\Bigg\{\Bigg(4\Big({b_{3L}^{U}}+{b_{3R}^{U}}\Big) \Delta_{3C}^{\alpha}-3b_{3C}^{U}\Big(\frac{5}{3}\Delta_{Y}^{\alpha}+\Delta_{2L}^{\alpha}\Big)\Bigg) {\large \ln}\left(\frac{M_{\alpha}}{M_Z}\right)\Bigg\}\nonumber \\
 &&\hspace*{+1.00cm}
 -\frac{1}{2\pi}\Bigg\{4\Big({b_{3L}^{U}}+{b_{3R}^{U}}\Big)b_{3C}^{2}-3b_{3C}^{U}\Big(\frac{5}{3}{b_{Y}^{2}}+{b_{2L}^{2}}\Big)\Bigg\} {\large \ln}\left(\frac{M_I}{M_Z}\right)\Bigg]
 \label{rel:alphaG}
\end{eqnarray}

\begin{eqnarray}
 &&\sin^{2}\theta_{W}= \frac{1}{4\Big({b_{3L}^{U}}+{b_{3R}^{U}}-2{b_{3C}^{U}} \Big)}
 \Bigg[\Big(3 b_{3L}^{U}-3 b_{3C}^{U}(1+\epsilon)\Big)- 4\Big\{b_{3L}^{U}(1-\epsilon)-b_{3R}^{U}(1+\epsilon)\Big\} \frac{\alpha^{-1}_{s}(M_Z)}{\alpha^{-1}_{em}(M_Z)}
 \nonumber \\
 &&\hspace*{+1.50cm}
 -\frac{\alpha_{em}(M_Z)}{16\pi}\sum_{\alpha=1,2}\Bigg\{ \Big[4\Big({b_{3L}^{U}}+{b_{3R}^{U}}-2{b_{3C}^{U}}\Big)\Big( \Delta_{2L}^{\alpha}(5-3\epsilon)-5\Delta_{Y}^{\alpha}(1+\epsilon)\Big)
 \nonumber \\
 &&\hspace*{+1.50cm}
 +4\Big(b^{U}_{3L}(1-\epsilon)-b^{U}_{3R}(1+\epsilon)\Big)\Big(8\Delta_{3C}^{\alpha}-3\Delta_{2L}^{\alpha}-5\Delta_{Y}^{\alpha}\Big)\Big]{\large \ln}\left(\frac{M_{\alpha}}{M_Z}\right)\Bigg\}+\nonumber \\
 &&\hspace*{+1.50cm}
 +\frac{\alpha_{em}(M_Z)}{16\pi}\Bigg\{4\Big({b_{3L}^{U}}+{b_{3R}^{U}}-2{b_{3C}^{U}}\Big) \Big(b^{2}_{2L}(5-3\epsilon)-5b^{2}_{Y} (1+\epsilon)\Big)
 \nonumber \\
 &&\hspace*{+1.50cm}
 +4\Big(b^{U}_{3L}(1-\epsilon)-b^{U}_{3R}(1+\epsilon)\Big)\Big(8b_{3C}^{2}-3b_{2L}^{2}-5b_{Y}^{2}\Big)\Bigg\}{\large \ln}\left(\frac{M_I}{M_Z}\right)
 \Bigg]
 \label{rel:sin}
\end{eqnarray}

Here $\Delta_{i(3C,2L,Y)}^{\alpha(1,2)}$ denote the increment of the corresponding beta coefficients due to addition of particles, such that $\Delta_{i}^{1}=(b_i^1-b_i)$ and $\Delta_{i}^{2}=(b_i^2-b_i^1)$. $\alpha_{em}^{-1}$ and $\alpha_{s}^{-1}$ are the inverse coupling constant for the electromagnetic interaction and strong interaction respectively. \\

It is observed that for a given choice of intermediate scales $M_I$ and  $M_{1,2}$ (mass of  additional particles), unification mass scale $M_U$ and the corresponding GUT coupling constant are unaffected by the gravitational correction parameter $\epsilon$. However $\sin^2\theta_W$ value is controlled by this  parameter. Thus for viable phenomenology, we fine tune  $\epsilon$ and the free mass parameters $M_1$, $M_2$ and $M_I$, such that $\sin^2\theta_W$ is in agreement with the accepted value $0.23129$ \cite{ParticleDataGroup:2018ovx}. Hence in our  numerical estimation, we strictly follow this constraint in order to obtain the unification mass $M_U$, the inverse GUT coupling constant $\alpha_{G}^{-1}$ and the proton lifetime $\tau_{p}$. Here we confine ourselves to the contribution from gauge dimension-6 operator for calculating the Proton lifetime$\left( p\rightarrow e^+\pi^0 \right)$ due to superheavy gauge boson exchange. By using dimensional analysis $\tau_{p}$ is given by
\begin{eqnarray}
 \tau_p=& C \frac{M_{U}^{4}}{m_{p}^{5}\alpha_{G}^{2}}
\end{eqnarray}

where $C$ is $\sim O(1)$ which contains all information about the flavor structure of this theory and $m_{p}$ is the mass of the proton. The experimental bound \cite{Super-Kamiokande:2016exg,Abe:2011ts,Yokoyama:2017mnt} on the lifetime for the above mentioned channel is as follows:
\begin{eqnarray}
& &\tau_p (p \to e^+ \pi^0) \big|_{SK}
> 1.6 \times 10^{34}\, \mbox{yrs}
\nonumber \\
& &\tau_p (p \to e^+ \pi^0) \big|_{HK, 2025} > 9.0 \times 10^{34}\, \mbox{yrs} \nonumber \\
& &\tau_p (p \to e^+ \pi^0) \big|_{HK, 2040} > 2.0 \times 10^{35}\, \mbox{yrs} 
\end{eqnarray}

In the next section we focus on the quantitative estimation for various viable models both non-SUSY and  SUSY inspired possibilities.

\section{Numerical Estimation for Unification mass, GUT coupling constant and Proton decay lifetime}
\label{sec:result}
It is noteworthy to mention that in $E_6$ GUT, with intermediate D-parity conserving trinification symmetry, the mass scale $M_I$ is shown to have vanishing multiloop, gravitational as well as threshold corrections. With D-parity violating models both in SUSY and non-SUSY cases in presence of gravitational correction, we can get a range of solutions for the free parameters $M_1$, $M_2$ and $M_I$ consistent with unification. However as mentioned before, it is observed that in the $E_6$ GUT, even with gravitational correction, to obtain phenomenologically consistent $M_U$ for comparatively low $M_I$, additional light particles are necessary. In the present section we analyze those possiblility with numerical estimation so as to have testable predictions. We refer to the Appendix for more details about the particle contents and the corresponding beta coefficients for the models. 

\subsection{NON-SUSY Models:}
It has been pointed out \cite{Chakrabortty:2017mgi} that, the non-SUSY $E_6$ GUT with $G_{333}$ intermediate symmetry (both D-parity conserving and broken) do not admit consistent unification with only $(1, \overline{3},3)$ of $27_H$. In presence of gravitational correction, it is possible, but with high $M_I$ closer to $M_U$. However to achieve comparatively low $M_I$ with consistent $M_U$, gravitational correction alone is not sufficient. The consistency of the present framework can be materialised by introducing additional particles either scalar or a fermion around TeV scale or higher. Four different models, which allow low $M_{I}\leq10^9$ GeV in comply with experimentally accessible proton decay, are analysed. The viability of the models is shown below with a quantitative estimation in Table-\ref{tab:non-susy}.

\begin{widetext}
\begin{center}
\begin{table}[h]
\centering
\vspace{-2pt}
\begin{tabular}{||c|c|c|c|c|c|c|c|c||}
\hline \hline
Models & Particles Content & Particle &Particle &$\epsilon$ &$M_{I}$ & ${M_U}$ & ${\alpha_G^{-1}}$ & $\tau_{p}$  \\
&$(M_I-M_U)$ & with mass&with mass & &(GeV) &(GeV) & &(years) \\
& &  $M_1$(GeV)&$M_{2}$(GeV) & & & & & \\

\hline
& &$\begin{array}{clcr}
                                 (8,1,0)_{S}
                                 \end{array}
$& & & & & &  \\
          & &$10^{5}$ & & $\begin{array}{clcr}
               -0.352147\\-0.188465
          \end{array}$& $\begin{array}{clcr}10^{6}\\10^{9}\end{array}$& $\begin{array}{clcr}10^{16.00}\\10^{16.05}\end{array}$ & $\begin{array}{clcr}33.1212\\35.3933\end{array}$  & $\begin{array}{clcr}3.15\times 10^{35}\\5.70\times 10^{35}\end{array}$ \\
      Model-I     &$\begin{array}{clcr}
                              27_{F} \oplus (1,\overline{3},3)_{{27}_{H}}\\
                              \left\{(1,8,8)
                              \oplus
                              (8,1,1)\right\}_{{650}_{H}}\end{array} $ & &$-$ & & & & &\\
           & & & & & & & &\\
          & & $10^{7}$& &$\begin{array}{clcr}
               -0.238286\\-0.186661
          \end{array}$& $\begin{array}{clcr}10^{8}\\10^{9}\end{array}$& $\begin{array}{clcr}10^{15.77}\\10^{15.78}\end{array}$ & $\begin{array}{clcr}34.9779\\35.7353\end{array}$  & $\begin{array}{clcr}4.22\times 10^{34}\\4.83\times 10^{34}\end{array}$ \\
          \hline
& &$\begin{array}{clcr}(1,3,0)_{F} \end{array}$ & $\begin{array}{clcr}(8,1,0)_{F} \end{array}$& & &  &   &  \\
          & & &$10^{5}$ &  $\begin{array}{clcr}
               -0.558954\\-0.418357
          \end{array}$& $\begin{array}{clcr}10^{6}\\10^{8}\end{array}$& $\begin{array}{clcr}10^{15.93}\\10^{16.10}\end{array}$ & $\begin{array}{clcr}29.0173\\30.6664\end{array}$  & $\begin{array}{clcr}1.27\times 10^{35}\\6.78\times 10^{35}\end{array}$ \\
  Model-II        &$\begin{array}{clcr}27_{F}\oplus
                              \left\{(1,8,1)\oplus(8,1,1)\right\}_{{78}_{F}}\\
                              (1,\overline{3},3)_{{27}_{H}}\oplus
                              (1,8,8)_{{650}_{H}}
                              \end{array}$ & $10^{3}$& & & & & &\\
          & & & & & & & &\\
          & & &$10^{6}$ &$\begin{array}{clcr}
               -0.476418\\-0.345651
          \end{array}$& $\begin{array}{clcr}10^{7}\\10^{9}\end{array}$& $\begin{array}{clcr}10^{15.75}\\10^{15.92}\end{array}$ & $\begin{array}{clcr}30.2816\\31.9307\end{array}$  & $\begin{array}{clcr}2.63\times 10^{34}\\1.40\times 10^{35}\end{array}$ \\
\hline
Model-III&$\begin{array}{clcr}27_{F}\oplus
                              \left\{(8,\overline{3},3)\oplus(1,6,\overline{6})\right\}_{351^{'}_H}\\
                              (1,\overline{3},3)_{{27}_{H}}\oplus
                              (1,8,8)_{{650}_{H}}
                              \end{array}$&$\begin{array}{clcr}(8,2,\frac{1}{2})_{S} \end{array}$ & $\begin{array}{clcr}(8,1,0)_{S} \end{array}$&  &   & & & \\
& &$10^{3}$ &$10^{4}$ & $\begin{array}{clcr}
               -0.831056
          \end{array}
$ &$\begin{array}{clcr}
              10^{9}
          \end{array}
$ &$\begin{array}{clcr}
              10^{16.22}
          \end{array}
$ &$\begin{array}{clcr}
               8.98183
          \end{array}
$ &$\begin{array}{clcr}
               1.76 \times 10^{35}
          \end{array}
$\\
\hline
Model-IV&$ \begin{array}{clcr}27_{F}\oplus(1,8,1)_{78_{F}}\oplus\\
                              (1,\overline{3},3)_{27_H}\oplus
                              \\\left\{(1,\overline{3},3)\oplus
                              (1,6,\overline{6})\right\}_{351^{'}_H}\end{array} $&$\begin{array}{clcr}(1,3,0)_{F} \end{array}$&$-$& &  & &   &  \\
& &$10^{3}$ & &$\begin{array}{clcr}
                -0.332393
          \end{array}$ & $\begin{array}{clcr}10^{8}\end{array}$&$\begin{array}{clcr}10^{15.82}\end{array}$  & $\begin{array}{clcr}38.2902\end{array}$& $\begin{array}{clcr} 8.02 \times 10^{34}\end{array}$\\
\hline \hline
\end{tabular}
\caption{Numerically estimated values for  $M_I$, $M_U$, $\alpha_G^{-1}$, $\tau_p$ for non-SUSY $E_6$ GUTs}.
\label{tab:non-susy}
\end{table} 
\end{center}
\end{widetext}

We now summarise the observations with relevant phenomenological implications. The Model-I utilises only $\left\{650_{H} \oplus 27_{H}\right\}$ with additional light multiplets $\left\{(1,8,8) \oplus (8,1,1)\right\}$ of $\{650_{H}\}$. The model can accommodate a color octet scalar, with possible mass $M_{1}$ ranging from $10^{5}-10^{7}$ GeV, in tune with $M_{I}$-$10^{6}-10^{9}$ GeV, such that the proton decay will be experimentally accessible in future. These color octet scalars are phenomenologically-rich example of physics beyond the SM. Model-II utilises $\left\{650_{H} \oplus27_{H}\right\}$ along with a non-standard fermion $78_{F}$, so as to allow a TeV scale electroweak triplet $(1,3,0)$, accompanied by a color octet $(8,1,0)$ fermion. The triplet fermion with even matter-parity may act as a stable dark matter candidate. The accompanied colored fermion may have mass ranging $10^{5}-10^{6}$ GeV, for a permissible $M_{I}$ within $10^{6}-10^{9}$ GeV. In Model-III we use the $\{351^{'}\}$ Higgs along with the conventional scalars $\{650_{H} \oplus 27_{H}\}$. Through gravitational correction we can have low $M_{I}$ with successful gauge unification. The most interesting feature of the model shows the presence of an electroweak color octet scalar \cite{Manohar:2006ga} $(8,2,\frac{1}{2})$ along with a $(8,1,0)$, which can be very informative from the collider physics point of view. In Model-IV we use the $\{351^{'}\}$ Higgs along with the conventional scalars $\{650_{H} \oplus 27_{H}\}$ and the $78_{F}$ for successful gauge unification. A TeV scale electroweak triplet fermion can be accommodated, which may be a stable dark matter candidate. The intermediate scale $M_{I}=10^{8}$ GeV, such that $M_{U}$ is consistent with respect to accessible proton decay. Model-I and II shows the Landau pole beyond the Planck scale. The Model-III is perturbative till $10^{16.82}$ GeV and Model-IV has Landau Pole position at $10^{17.95}$ GeV.

\subsection{SUSY Models:}

SUSY $E_{6}$ models with D-parity conserving $G_{333}$ intermediate symmetry have been discussed by some authors \cite{Chakrabortty:2017mgi,Dash:2019bdh} which accommodate high intermediate scale $>$ $10^{15}$ GeV. However gravity induced D-parity broken models can trigger low $M_{I}$ scale as low as TeV scale. In the present framework we observe that low $M_{I}$ may not always comply with experimentally accessible proton decay constraint. Additional light multiplets are required to solve the purpose as has been mentioned in case of non-SUSY models. We now consider three different models to analyse the phenomenological viability. For simplicity, we assume that the SUSY scale $M_{S}$ coincides  with $M_{Z}$ in all models. The numerical estimation is given in Table-\ref{tab:susy}.

\begin{widetext}
\begin{center}
\begin{table}[h]
\centering
\vspace{-2pt}
\begin{tabular}{||c|c|c|c|c|c|c|c|c||}
\hline \hline
Models & Particles Content & Particle &Particle &$\epsilon$ &$M_{I}$ & ${M_U}$ & ${\alpha_G^{-1}}$ & $\tau_{p}$  \\
&$(M_I-M_U)$ & with mass&with mass & &(GeV) &(GeV) & &(years) \\
& &  $M_1$(GeV)&$M_{2}$(GeV) & & & & & \\
\hline
Model-I &$ \begin{array}{clcr}27_{F}\oplus
                              2\left\{(1,\overline{3},3)_{27_{H}}\oplus(1,3,\overline{3})_{\overline{27}_H}\right\}\\\oplus
                               (1,1,8)_{650_{H}}\end{array} $&$-
$&$-$ &$\begin{array}{clcr}-0.839354\\-0.699153\\-0.582832\\-0.48477\end{array} $& $\begin{array}{clcr}10^{4}\\10^{5}\\10^{6}\\10^{7}\end{array}$ &$\begin{array}{clcr}10^{16.34}\\10^{16.34}\\10^{16.34}\\10^{16.34}\end{array}$&$\begin{array}{clcr}10.7103\\11.8097\\12.9091\\14.0085\end{array}$ & $\begin{array}{clcr}7.54 \times 10^{35}\\9.17 \times 10^{35}\\1.10 \times 10^{36}\\1.29 \times 10^{36}\end{array}$ \\
\hline
Model-II &$ \begin{array}{clcr}27_{F} \oplus\left\{(1,8,1) \oplus (1,1,8)\right \}_{78_F}\\\oplus
(1,\overline{3},3)_{27_{H}}\oplus (1,3,\overline{3})_{\overline{27}_H}\\\oplus
                              
(1,1,8)_{650_{H}} \end{array} $&$\begin{array}{clcr}(1,3,0)_{F}\\10^{3} \end{array}$&$-$ &$\begin{array}{clcr}-0.887257\\-0.78604\\-0.702064\\-0.631268\end{array} $& $\begin{array}{clcr}10^{4}\\10^{5}\\10^{6}\\10^{7}\end{array}$ &$\begin{array}{clcr}10^{16.24}\\10^{16.14}\\10^{16.04}\\10^{15.94}\end{array}$&$\begin{array}{clcr}10.7103\\11.8097\\12.9091\\14.0085\end{array}$ & $\begin{array}{clcr}3.00 \times 10^{35}\\1.45 \times 10^{35}\\6.91 \times 10^{34}\\3.24 \times 10^{34}\end{array}$ \\
\hline
& &$\begin{array}{clcr}(8,1,0)_{S} \end{array}$ & 
                                                                                                                        
                                                                        & &  & & &\\
                                                                        Model-III&$\begin{array}{clcr}27_{F}\oplus(1,\overline{3},3)_{27_{H}}\\\oplus(1,6,\overline{6})_{351_{H}^{'}}\oplus(8,1,1)_{650_{H}}
                               \end{array} $ & &$-$ & & & & &\\
& & $\begin{array}{clcr}10^{6}\\10^{7}\\10^{8} \end{array}$& &  $\begin{array}{clcr}
        -0.55056\\-0.450654\\-0.356779
          \end{array}
$& $\begin{array}{clcr}
        10^{12.45}\\10^{12.65}\\10^{13}
          \end{array}
$  &$\begin{array}{clcr}
        10^{16.04}\\10^{15.95}\\10^{15.96}
          \end{array}
$&$\begin{array}{clcr}
        8.95951\\10.3765\\11.8485
          \end{array}
$&$\begin{array}{clcr}
        3.33 \times 10^{34}\\1.95 \times 10^{34}\\2.79 \times 10^{34}
          \end{array}
$\\
\hline \hline
\end{tabular}
\caption{Numerically estimated values for  $M_I$, $M_U$, $\alpha_G^{-1}$ and $\tau_p$ for SUSY $E_6$ GUTs}.
\label{tab:susy}
\end{table} 
\end{center}
\end{widetext}

We now discuss the qualitative implication of the SUSY models. The Model-I uses the $\{650_{H} \oplus 2(27_{H} \oplus \overline{27}_{H})\}$ for successful gauge unification, where we introduce additional light multiplet $(1,1,8)$ of $\{650\}$ at $M_{I}$. The Model ensures gauge unification at a fixed point i.e. $M_{U}=10^{16.34}$ GeV for all possible values of $M_{I}$, making it more predictive. However in order to satisfy the proton decay check point, the intermediate scale $M_{I}$ is constrained to take a value within  $10^{4}-10^{8}$ GeV. The Model-II utilises the $\{650_{H} \oplus (27_{H} \oplus \overline{27}_{H})\}$ and a non-standard fermion $78_{F}$ to achieve gauge unification within the accessible proton decay for low $M_{I}$ in the range $10^{4}-10^{7}$ GeV. The Model accommodates a TeV scale electroweak triplet, which may be a dark matter candidate. The Model-III has the scalar contents $\{650_{H} \oplus 27_{H} \oplus {351}^{'}_{H}\}$ which includes a color octet $(8,1,0)$ scalar with permissible mass $10^{6}-10^{8}$ GeV so as to comply with the proton decay. However, it is observed that the high intermediate scale $M_{I}$ is inevitable in presence of color octet scalar unlike the case of non-SUSY model.
  In Models I and II, we note that for given $M_{I}$, the inverse gut couplings are same, even if the models are distinct. Accordingly, we quote an interesting remark for the SUSY GUT models with $G_{333}$ intermediate symmetry, It states that:
 \textit{\textquotedblleft{For all $E_6$ supersymmetric Grand Unified Theories (GUTs) with intermediate D-parity violating trinification symmetry, in presence of the conventional MSSM (minimal SUSY standard model) particles along with or without additional color singlets at the lower scale (within $M_{Z}-M_{I}$) the inverse GUT coupling attains a constant value corresponding to a given $M_{I}$.}\textquotedblright}
  It is estimated to be,
  \begin{eqnarray}
   \alpha_{G}^{-1}=\alpha_{s}^{-1}+\frac{3}{2\pi} {\large \ln}\left(\frac{M_I}{M_Z}\right)
  \end{eqnarray}

Thus $\alpha_{G}^{-1}$ solely depends on $M_{I}$, leading to a constant value for all such models. However the above remark is valid with one-loop renormalisation effects only.\\

As has been mentioned before all the above models are perturbative till a point slightly above the gauge unification scale $M_{U}$. This is due to large particle contents leading to high contribution to the corresponding beta coefficient. The early non-perturbative region \cite{Babu:2015psa} is mainly due to the large representation $\{650\}$, which is unavoidable in case of trinification symmetry at the intermediate scale as well as for inclusion of gravitational correction.

\vspace*{0.95cm}
\section{Conclusion}
\label{sec:conclusion}

 We have investigated the effect of gravitational correction through non-renormalizable operators in $E_{6}$ GUT models (with and without SUSY) with the main objective to achieve low intermediate scale $M_{I}$. The analysis is confined to D-parity violating trinification symmetry at the intermediate scale. It is nice to mention that it is indeed possible to achieve low $M_{I}\leq10^{9}$ GeV for the models with successful gauge unification in comply with the verifiable proton lifetime. As far as phenomenological implication is concerned, a TeV scale color octet electroweak doublet $(8,2,\frac{1}{2})$ \cite{Manohar:2006ga} scalar can be predicted in $E_{6}$ GUT (non-SUSY Model-III), which can be very interesting from the collider physics point of view. These colored scalars, would be produced in pairs with a large rate at the LHC, which may affect the Higgs boson production cross section in gluon fusion \cite{Hayreter:2017wra,Cheng:2018mkc}. Based on the predictions SUSY inspired models-I and II, we establish a very interesting remark for $E_{6}$ SUSY GUTs with trinification $G_{333}(g_{3L}\neq g_{3R})$ symmetry at the intermediate scale $M_{I}$, such that $\alpha_{G}^{-1}$ remains constant for a given $M_{I}$, irrespective of the models chosen, even in presence of additional particles. The models also give interesting predictions on dark matter candidates $(1,3,0)_{F}$ and occurrence of color octet scalar $(8,1,0)$. These GUT models with low $M_{I}$ may contribute to matter asymmetry through non-thermal or resonant leptogenesis, thermal leptogenesis without gravitino overabundance. However the SUSY models suffer from obvious problem of non-perturbativity beyond the unification scale. We hope the problem can be solved with the concept of Seiberg duality \cite{Abel:2008tx,Abel:2009bj}.

 With reference to the cosmological issues \cite{Kibble:1982ae,Lazarides:2019xai,Lazarides:2021tua}, since the discrete D-parity is broken at the GUT scale, the models are not plagued by the generation of topological defects like strings or domain-walls bounded by strings. The subsequent breaking to the standard model may or may not generate any monopoles depending on the Higgs sector used. It is worth noting that the monopoles associated with the spontaneous breaking of the gauge symmetry $SU(3)_{L} \otimes SU(3)_{R}\rightarrow SU(2)_{L} \otimes U(1)_{Y}$ may have low-lying masses, which will be interesting to investigate from a cosmological perspective.

\noindent
 \vspace*{-1.19cm}
\section*{Acknowledgments}
\vspace*{-0.50cm}
Chandini Dash is grateful to the Department 
of Science and Technology, Govt. of India for INSPIRE Fellowship/2015/IF150787 in support of her research work. 
\vspace*{+4.50cm}
%%%%%%%%%%%%%%%%%%%%%%%%%%%%%
\appendix
%%%%%%%%%%%%%%%%%%%%%%%%%%%%%
\section{Fields and one-loop beta coefficients for different Models}
\subsection{NON-SUSY CASES}
\begin{table}[h!]
\begin{tabular}{|c|c|c|c|c|}
\hline
Group & Range  & Higgs & Fermions & beta coefficients \\

 &  of masses & content & content  &  \\
\hline
$G_{321}$ & $M_{Z}-M_{1}$ & $ \begin{array}{clcr}
                                (1,2,-\frac{1}{2})_{27} 
                                 \end{array} $ &SMPs &$\begin{pmatrix}{b_{3C}}=-7\\
                                 {b_{2L}}=-\frac{19}{6}\\
                                 {b_{Y}}=\frac{41}{10}\end{pmatrix}$\\
\hline
$G_{321}$ & $M_{1}-M_{I}$ & $ \begin{array}{clcr}
                                (1,2,-\frac{1}{2})_{27} \\  (8,1,0)_{650}
                                
                                 \end{array} $ &SMPs &$\begin{pmatrix}{b^{1}_{3C}}=-6\\
                                 {b^{1}_{2L}}=-\frac{19}{6}\\
                                 {b^{1}_{Y}}=\frac{41}{10}\end{pmatrix}$\\
\hline
$G_{333}$ & $M_{I}-M_{U}$ & $ \begin{array}{clcr}
                              (1,\overline{3},3)_{27}\\
                              \left\{(1,8,8)
                              \oplus
                              (8,1,1)\right\}_{650}\end{array} $ & $27$ &$\begin{pmatrix} {b_{3C}^{U}}=-4 \\ {b_{3L}^{U}}=\frac{7}{2} \\ {b_{3R}^{U}}=\frac{7}{2} \end{pmatrix}$\\
\hline
\end{tabular}
\caption{\pmb{ Model-I}}
\label{tab:Model-I}
\end{table}

\begin{table}[h!]
\begin{tabular}{|c|c|c|c|c|}
\hline
Group & Range  & Higgs & Fermions & beta coefficients \\

 &  of masses & content & content  &  \\
\hline
$G_{321}$ & $M_{Z}-M_{1}$ & $ \begin{array}{clcr}
                                (1,2,-\frac{1}{2})_{27} 
                                 \end{array} $ &SMPs &$\begin{pmatrix}{b_{3C}}=-7\\
                                 {b_{2L}}=-\frac{19}{6}\\
                                 {b_{Y}}=\frac{41}{10}\end{pmatrix}$\\
\hline
$G_{321}$ & $M_{1}-M_{2}$ & $(1,2,-\frac{1}{2})_{27}$ &  SMPs$+ (1,3,0)_{78}$&$\begin{pmatrix}{b_{3C}^{1}}=-7\\
                                 {b_{2L}^{1}}=-\frac{11}{6}\\
                                 {b_{Y}^{1}}=\frac{41}{10}\end{pmatrix}$\\
\hline
$G_{321}$ & $M_{2}-M_{I}$ & $(1,2,-\frac{1}{2})_{27}$ & SMPs$+ \left\{(1,3,0)\oplus(8,1,0)\right\}_{78}
                                  $ &$\begin{pmatrix}{b_{3C}^{2}}=-5\\
                                 {b_{2L}^{2}}=-\frac{11}{6}\\
                                 {b_{Y}^{2}}=\frac{41}{10}\end{pmatrix}$\\
\hline
$G_{333}$ & $M_{I}-M_{U}$ & $ \begin{array}{clcr}
                              (1,\overline{3},3)_{27}\\
                              (1,8,8)_{650}
                              \end{array} $ & $ 27\oplus
                              \left\{(1,8,1)\oplus(8,1,1)\right\}_{78}$ &$\begin{pmatrix} {b_{3C}^{U}}=-3 \\ {b_{3L}^{U}}=\frac{11}{2} \\ {b_{3R}^{U}}=\frac{7}{2} \end{pmatrix}$\\
\hline
\end{tabular}
\caption{\pmb{Model-II}}
\label{tab:Model-II}
\end{table}
\begin{table}[h!]
\begin{tabular}{|c|c|c|c|c|}
\hline
Group & Range  & Higgs & Fermions & beta coefficients \\

 &  of masses & content & content  &  \\
\hline
$G_{321}$ & $M_{Z}-M_{1}$ & $ \begin{array}{clcr}
                                (1,2,-\frac{1}{2})_{27} 
                                 \end{array} $ &SMPs &$\begin{pmatrix}{b_{3C}}=-7\\
                                 {b_{2L}}=-\frac{19}{6}\\
                                 {b_{Y}}=\frac{41}{10}\end{pmatrix}$\\
\hline
$G_{321}$ & $M_{1}-M_{2}$ & $\begin{array}{clcr}(1,2,-\frac{1}{2})_{27} \\ (8,2,\frac{1}{2})_{351{'}}\end{array}$ &SMPs&$\begin{pmatrix}{b_{3C}^{1}}=-5\\
                                 {b_{2L}^{1}}=-\frac{11}{6}\\
                                 {b_{Y}^{1}}=\frac{49}{10}\end{pmatrix}$\\
\hline
$G_{321}$ & $M_{2}-M_{I}$ & $\begin{array}{clcr}(1,2,-\frac{1}{2})_{27}\\  \{(8,2,\frac{1}{2})\oplus(8,1,0)\}_{351^{'}}
                             \end{array}$ &SMPs &$\begin{pmatrix}{b_{3C}^{2}}=-4\\
                                 {b_{2L}^{2}}=-\frac{11}{6}\\
                                 {b_{Y}^{2}}=\frac{49}{10}\end{pmatrix}$\\
\hline
$G_{333}$ & $M_{I}-M_{U}$ & $ \begin{array}{clcr}
                              (1,\overline{3},3)_{27}\\
                              (1,8,8)_{650}\\
                              \{(1,6,\overline{6}) \oplus (8,\overline{3},3)\}_{351^{'}}
                              \end{array} $ & $27$ &$\begin{pmatrix} {b_{3C}^{U}}=4 \\ {b_{3L}^{U}}=\frac{25}{2} \\ {b_{3R}^{U}}=\frac{25}{2} \end{pmatrix}$\\
\hline
\end{tabular}
\caption{\pmb{Model-III}}.
\label{tab:oneloopresults}
\end{table}
\begin{table}[h!]
\begin{tabular}{|c|c|c|c|c|}
\hline
Group & Range  & Higgs & Fermions & beta coefficients \\

 &  of masses & content & content  &  \\
\hline
$G_{321}$ & $M_{Z}-M_{1}$ & $ \begin{array}{clcr}
                                (1,2,-\frac{1}{2})_{27} 
                                 \end{array} $ &SMPs &$\begin{pmatrix}{b_{3C}}=-7\\
                                 {b_{2L}}=-\frac{19}{6}\\
                                 {b_{Y}}=\frac{41}{10}\end{pmatrix}$\\
\hline
$G_{321}$ & $M_{1}-M_{I}$ & $ \begin{array}{clcr}
                                (1,2,-\frac{1}{2})_{27}
                                
                                 \end{array} $ &SMPs$+(1,3,0)_{78}$ &$\begin{pmatrix}{b^{1}_{3C}}=-7\\
                                 {b^{1}_{2L}}=-\frac{11}{6}\\
                                 {b^{1}_{Y}}=\frac{41}{10}\end{pmatrix}$\\
\hline
$G_{333}$ & $M_{I}-M_{U}$ & $ \begin{array}{clcr}
                              (1,\overline{3},3)_{27}
                              \\\left\{(1,\overline{3},3)\oplus
                              (1,6,\overline{6})\right\}_{351^{'}}\end{array} $ & $ 27\oplus(1,8,1)_{78}$ &$\begin{pmatrix} {b_{3C}^{U}}=-5 \\ {b_{3L}^{U}}=3 \\ {b_{3R}^{U}}=1 \end{pmatrix}$\\
\hline
\end{tabular}
\caption{\pmb{Model-IV}}
\label{tab:Model-IV}
\end{table}

 \vspace*{+1.0cm}
\subsection{SUSY CASES}

% \subsubsection{\pmb{CASE-I}}
%  \vspace*{-2.5cm}
\begin{table}[h!]
\begin{tabular}{|c|c|c|c|c|}
\hline
Group & Range  & Higgs & Fermions & beta coefficients \\

 &  of masses & content & content  &  \\
\hline
$G_{321}$ & $M_{Z}-M_{I}$ & $ \begin{array}{clcr}
                                (1,2,\pm\frac{1}{2})_{{27}\oplus{\overline{27}}} 
                                 \end{array} $ &SMPs &$\begin{pmatrix}{b_{3C}}=-3\\
                                 {b_{2L}}=1\\
                                 {b_{Y}}=\frac{33}{5}\end{pmatrix}$\\
\hline
$G_{333}$ & $M_{I}-M_{U}$ & $ \begin{array}{clcr}
                              \{(1,\overline{3},3)\oplus(1,3,\overline{3})\}_{27\oplus \overline{27}}\\
                              \{(1,\overline{3},3)\oplus(1,3,\overline{3})\}_{27\oplus \overline{27}}\\
                               (1,1,8)_{650}\end{array} $ & $ 27$ &$\begin{pmatrix} {b_{3C}^{U}}=0 \\ {b_{3L}^{U}}=6 \\ {b_{3R}^{U}}=9 \end{pmatrix}$\\
\hline
\end{tabular}
\caption{\pmb{Model-I}}.
\label{tab:Model-Is}
\end{table}

\begin{table}[h!]
\begin{tabular}{|c|c|c|c|c|}
\hline
Group & Range  & Higgs & Fermions & one-loop   
 \\

 &  of masses & content & content  & beta coefficients  \\
\hline
$G_{321}$ & $M_{Z}-M_{1}$ & $ (1,2,\pm\frac{1}{2})_{{27}\oplus{\overline{27}}}$& SMPs  &$\begin{pmatrix}{b_{3C}}=-3\\
{b_{2L}}=1\\
{b_{Y}}=\frac{33}{5}\end{pmatrix}$\\
\hline
$G_{321}$ & $M_{1}-M_{I}$ & $ \begin{array}{clcr} (1,2,\pm\frac{1}{2})_{{27}\oplus{\overline{27}}}\end{array}$& SMPs $+ (1,3,0)_{78}$ &$\begin{pmatrix}{b_{3C}^{1}}=-3\\
{b_{2L}^{1}}=3\\
{b_{Y}^{1}}=\frac{33}{5}\end{pmatrix}$\\
\hline
$G_{333}$ & $M_{I}-M_{U}$ & $ \begin{array}{clcr}
\{(1,\overline{3},3)\oplus(1,3,\overline{3})\}_{27\oplus \overline{27}}\\
                              
(1,1,8)_{650} \end{array} $& $27 \oplus\left\{(1,8,1) \oplus (1,1,8)\right \}_{78}$  &$\begin{pmatrix} {b_{3C}^{U}}=0 \\ {b_{3L}^{U}}=6 \\ {b_{3R}^{U}}=9 \end{pmatrix}$ \\
\hline
\end{tabular}
\caption{\pmb{Model-II}}.
\label{tab:Model-II}
\end{table}

\begin{table}[h!]
\begin{tabular}{|c|c|c|c|c|}
\hline
Group & Range  & Higgs & Fermions & one-loop   
 \\

 &  of masses & content & content  & beta coefficients  \\
\hline
$G_{321}$ & $M_{Z}-M_{1}$ & $ (1,2,\pm\frac{1}{2})_{{27}}$ & SMPs &$\begin{pmatrix}{b_{3C}}=-3\\
{b_{2L}}=1\\
{b_{Y}}=\frac{33}{5}\end{pmatrix}$\\
\hline
$G_{321}$ & $M_{1}-M_{I}$ & $ \begin{array}{clcr} (1,2,\pm\frac{1}{2})_{27}\oplus (8,1,0)_{650}\end{array}$& SMPs &$\begin{pmatrix}{b_{3C}^{\prime}}=0\\
{b_{2L}^{\prime}}=1\\
{b_{Y}^{\prime}}=\frac{33}{5}\end{pmatrix}$\\
\hline
$G_{333}$ & $M_{I}-M_{U}$ & $ \begin{array}{clcr}
(1,\overline{3},3)_{27}\\(1,6,\overline{6})_{351^{'}}\\
 (8,1,1)_{650}
\end{array} $ & $27$ &$\begin{pmatrix} {b_{3C}^{''}}=3 \\ {b_{3L}^{''}}=\frac{33}{2} \\ {b_{3R}^{''}}=\frac{33}{2} \end{pmatrix}$ \\
\hline
\end{tabular}
\caption{\pmb{Model-III}}.
\label{tab:Model-III}
\end{table}
where SMPs denotes the standard model particles.
\vspace*{+4.0cm}
%\clearpageon 
% \bibliographystyle{utphys.bst}
\bibliographystyle{utcaps_mod}
\bibliography{E62}

\providecommand{\href}[2]{#2}\begingroup\raggedright\begin{thebibliography}{10}

\bibitem{Babu:1985gi}
K.~S. Babu, X.-G. He, and S.~Pakvasa, ``{\em {Neutrino Masses and Proton Decay
  Modes in $SU(3) \times SU(3) \times SU(3)$ Trinification}},''
  \href{http://dx.doi.org/10.1103/PhysRevD.33.763}{Phys. Rev. D {\normalfont
  \bfseries 33} (1986)  763}.

\bibitem{Nishimura:1988fp}
H.~Nishimura and A.~Okunishi, ``{\em {STRONG CP PROBLEM AND NUCLEON STABILITY
  IN $SU(3) \times SU(3) \times SU(3)$ TRINIFICATION MODEL}},''
  \href{http://dx.doi.org/10.1016/0370-2693(88)90951-3}{Phys. Lett. B
  {\normalfont \bfseries 209} (1988)  307--310}.

\bibitem{Sayre:2006ma}
J.~Sayre, S.~Wiesenfeldt, and S.~Willenbrock, ``{\em {Minimal
  trinification}},'' \href{http://dx.doi.org/10.1103/PhysRevD.73.035013}{Phys.
  Rev. D {\normalfont \bfseries 73} (2006)  035013},
  \href{http://arxiv.org/abs/hep-ph/0601040}{{\normalfont \ttfamily
  arXiv:hep-ph/0601040}}.

\bibitem{Cauet:2010ng}
C.~Cauet, H.~Pas, S.~Wiesenfeldt, H.~Pas, and S.~Wiesenfeldt, ``{\em
  {Trinification, the Hierarchy Problem and Inverse Seesaw Neutrino Masses}},''
  \href{http://dx.doi.org/10.1103/PhysRevD.83.093008}{Phys. Rev. D {\normalfont
  \bfseries 83} (2011)  093008},
  \href{http://arxiv.org/abs/1012.4083}{{\normalfont \ttfamily
  arXiv:1012.4083}}.

\bibitem{Stech:2014tla}
B.~Stech, ``{\em {Trinification Phenomenology and the structure of Higgs
  Bosons}},'' \href{http://dx.doi.org/10.1007/JHEP08(2014)139}{JHEP
  {\normalfont \bfseries 08} (2014)  139},
  \href{http://arxiv.org/abs/1403.2714}{{\normalfont \ttfamily
  arXiv:1403.2714}}.

\bibitem{Hetzel:2015cca}
J.~Hetzel, \href{http://dx.doi.org/10.11588/heidok.00018259}{{\em
  {Phenomenology of a left-right-symmetric model inspired by the trinification
  model}}}.
\newblock PhD thesis, Inst. Appl. Math., Heidelberg, 2015.
\newblock \href{http://arxiv.org/abs/1504.06739}{{\normalfont \ttfamily
  arXiv:1504.06739}}.

\bibitem{Babu:2017xlu}
K.~S. Babu, B.~Bajc, M.~Nemev\v{s}ek, and Z.~Tavartkiladze, ``{\em
  {Trinification at the TeV scale}},''
  \href{http://dx.doi.org/10.1063/1.5010106}{AIP Conf. Proc. {\normalfont
  \bfseries 1900} (2017) no.~1, 020002}.

\bibitem{Gursey:1975ki}
F.~Gursey, P.~Ramond, and P.~Sikivie, ``{\em {A Universal Gauge Theory Model
  Based on $E_6$}},''
\href{http://dx.doi.org/10.1016/0370-2693(76)90417-2}{Phys. Lett. {\normalfont
  \bfseries 60B} (1976)  177--180}.
%%CITATION = PHLTA,60B,177;%%.

\bibitem{Shafi:1978gg}
Q.~Shafi, ``{\em {$E_6$ as a Unifying Gauge Symmetry}},''
  \href{http://dx.doi.org/10.1016/0370-2693(78)90248-4}{Phys. Lett. B
  {\normalfont \bfseries 79} (1978)  301--303}.

\bibitem{Chang:1983fu}
D.~Chang, R.~N. Mohapatra, and M.~K. Parida, ``{\em {Decoupling Parity and
  $SU(2)_R$ Breaking Scales: A New Approach to Left-Right Symmetric Models}},''
\href{http://dx.doi.org/10.1103/PhysRevLett.52.1072}{Phys. Rev. Lett.
  {\normalfont \bfseries 52} (1984)  1072}.
%%CITATION = PRLTA,52,1072;%%.

\bibitem{Chang:1984uy}
D.~Chang, R.~N. Mohapatra, and M.~K. Parida, ``{\em {A New Approach to
  Left-Right Symmetry Breaking in Unified Gauge Theories}},''
\href{http://dx.doi.org/10.1103/PhysRevD.30.1052}{Phys. Rev. {\normalfont
  \bfseries D30} (1984)  1052}.
%%CITATION = PHRVA,D30,1052;%%.

\bibitem{Stech:2003sb}
B.~Stech and Z.~Tavartkiladze, ``{\em {Fermion masses and coupling unification
  in $E_6$: Life in the desert}},''
  \href{http://dx.doi.org/10.1103/PhysRevD.70.035002}{Phys. Rev. D {\normalfont
  \bfseries 70} (2004)  035002},
  \href{http://arxiv.org/abs/hep-ph/0311161}{{\normalfont \ttfamily
  arXiv:hep-ph/0311161}}.

\bibitem{Wang:2011eh}
F.~Wang, ``{\em {Supersymmetry Breaking Scalar Masses and Trilinear Soft Terms
  From High-Dimensional Operators in $E_6$ SUSY GUT}},''
  \href{http://dx.doi.org/10.1016/j.nuclphysb.2011.05.017}{Nucl. Phys. B
  {\normalfont \bfseries 851} (2011)  104--142},
  \href{http://arxiv.org/abs/1103.0069}{{\normalfont \ttfamily
  arXiv:1103.0069}}.

\bibitem{Chakrabortty:2017mgi}
J.~Chakrabortty, R.~Maji, S.~K. Patra, T.~Srivastava, and S.~Mohanty, ``{\em
  {Roadmap of left-right models based on GUTs}},''
  \href{http://dx.doi.org/10.1103/PhysRevD.97.095010}{Phys. Rev. D {\normalfont
  \bfseries 97} (2018) no.~9, 095010},
  \href{http://arxiv.org/abs/1711.11391}{{\normalfont \ttfamily
  arXiv:1711.11391}}.

\bibitem{Dash:2019bdh}
C.~Dash, S.~Mishra, and S.~Patra, ``{\em {Theorem on vanishing contributions to
  $\sin^2\theta_W$ and intermediate mass scale in Grand Unified Theories with
  trinification symmetry}},''
  \href{http://dx.doi.org/10.1103/PhysRevD.101.055039}{Phys. Rev. D
  {\normalfont \bfseries 101} (2020) no.~5, 055039},
  \href{http://arxiv.org/abs/1911.11528}{{\normalfont \ttfamily
  arXiv:1911.11528}}.

\bibitem{Dash:2020jlc}
C.~Dash, S.~Mishra, S.~Patra, and P.~Sahu, ``{\em {Threshold effects on
  prediction for proton decay in non-supersymmetric $E_6$ GUT with intermediate
  trinification symmetry}},''
  \href{http://dx.doi.org/10.1016/j.nuclphysb.2020.115239}{Nucl. Phys. B
  {\normalfont \bfseries 962} (2021)  115239},
  \href{http://arxiv.org/abs/2004.14188}{{\normalfont \ttfamily
  arXiv:2004.14188}}.

\bibitem{Shafi:1983gz}
Q.~Shafi and C.~Wetterich, ``{\em {Modification of {GUT} Predictions in the
  Presence of Spontaneous Compactification}},''
\href{http://dx.doi.org/10.1103/PhysRevLett.52.875}{Phys. Rev. Lett.
  {\normalfont \bfseries 52} (1984)  875}.
%%CITATION = PRLTA,52,875;%%.

\bibitem{Hill:1983xh}
C.~T. Hill, ``{\em {Are There Significant Gravitational Corrections to the
  Unification Scale?}},''
\href{http://dx.doi.org/10.1016/0370-2693(84)90451-9}{Phys. Lett. {\normalfont
  \bfseries 135B} (1984)  47--51}.
%%CITATION = PHLTA,135B,47;%%.

\bibitem{Rizzo:1984mk}
T.~G. Rizzo, ``{\em {Gravitational Corrections to the Unification Scale in
  SO(10) With a Low Right-handed Mass Scale}},''
  \href{http://dx.doi.org/10.1016/0370-2693(84)91254-1}{Phys. Lett. B
  {\normalfont \bfseries 142} (1984)  163--167}.

\bibitem{Patra:1991dy}
P.~K. Patra and M.~K. Parida, ``{\em {Spontaneous compactification effects on
  $SO(10)$ grand unification with $SU(2)_{L} \times SU(2)_{R} \times SU(4)_{C}$
  intermediate symmetry}},''
  \href{http://dx.doi.org/10.1103/PhysRevD.44.2179}{Phys. Rev. D {\normalfont
  \bfseries 44} (1991)  2179--2187}.

\bibitem{Chakrabortty:2008zk}
J.~Chakrabortty and A.~Raychaudhuri, ``{\em {A Note on dimension-5 operators in
  GUTs and their impact}},''
  \href{http://dx.doi.org/10.1016/j.physletb.2009.01.065}{Phys. Lett. B
  {\normalfont \bfseries 673} (2009)  57--62},
  \href{http://arxiv.org/abs/0812.2783}{{\normalfont \ttfamily
  arXiv:0812.2783}}.

\bibitem{Huang:2017uli}
C.-S. Huang, W.-J. Li, and X.-H. Wu, ``{\em {$E_6$ GUT through effects of
  dimension-5 operators}},''
  \href{http://dx.doi.org/10.1088/2399-6528/aa98d1}{J. Phys. Comm. {\normalfont
  \bfseries 1} (2017) no.~5, 055025},
  \href{http://arxiv.org/abs/1705.01411}{{\normalfont \ttfamily
  arXiv:1705.01411}}.

\bibitem{Chang:2004pb}
D.~Chang, T.~Fukuyama, Y.-Y. Keum, T.~Kikuchi, and N.~Okada, ``{\em
  {Perturbative $SO(10)$ grand unification}},''
  \href{http://dx.doi.org/10.1103/PhysRevD.71.095002}{Phys. Rev. D {\normalfont
  \bfseries 71} (2005)  095002},
  \href{http://arxiv.org/abs/hep-ph/0412011}{{\normalfont \ttfamily
  arXiv:hep-ph/0412011}}.

\bibitem{Kopp:2009xt}
J.~Kopp, M.~Lindner, V.~Niro, and T.~E.~J. Underwood, ``{\em {On the
  Consistency of Perturbativity and Gauge Coupling Unification}},''
  \href{http://dx.doi.org/10.1103/PhysRevD.81.025008}{Phys. Rev. D {\normalfont
  \bfseries 81} (2010)  025008},
  \href{http://arxiv.org/abs/0909.2653}{{\normalfont \ttfamily
  arXiv:0909.2653}}.

\bibitem{ParticleDataGroup:2018ovx}
{\normalfont \bfseries Particle Data Group}, M.~Tanabashi {\em et al.}, ``{\em
  {Review of Particle Physics}},''
  \href{http://dx.doi.org/10.1103/PhysRevD.98.030001}{Phys. Rev. D {\normalfont
  \bfseries 98} (2018) no.~3, 030001}.

\bibitem{Super-Kamiokande:2016exg}
{\normalfont \bfseries Super-Kamiokande}, K.~Abe {\em et al.}, ``{\em {Search
  for proton decay via $p \to e^+\pi^0$ and $p \to \mu^+\pi^0$ in 0.31
  megaton\textperiodcentered{}years exposure of the Super-Kamiokande water
  Cherenkov detector}},''
  \href{http://dx.doi.org/10.1103/PhysRevD.95.012004}{Phys. Rev. D {\normalfont
  \bfseries 95} (2017) no.~1, 012004},
  \href{http://arxiv.org/abs/1610.03597}{{\normalfont \ttfamily
  arXiv:1610.03597}}.

\bibitem{Abe:2011ts}
K.~Abe {\em et al.}, ``{\em {Letter of Intent: The Hyper-Kamiokande Experiment
  --- Detector Design and Physics Potential ---}},''
  \href{http://arxiv.org/abs/1109.3262}{{\normalfont \ttfamily
  arXiv:1109.3262}}.

\bibitem{Yokoyama:2017mnt}
{\normalfont \bfseries Hyper-Kamiokande Proto}, M.~Yokoyama, ``{\em {The
  Hyper-Kamiokande Experiment}},'' in {\em {Prospects in Neutrino Physics}}.
\newblock 4, 2017.
\newblock \href{http://arxiv.org/abs/1705.00306}{{\normalfont \ttfamily
  arXiv:1705.00306}}.

\bibitem{Manohar:2006ga}
A.~V. Manohar and M.~B. Wise, ``{\em {Flavor changing neutral currents, an
  extended scalar sector, and the Higgs production rate at the CERN LHC}},''
  \href{http://dx.doi.org/10.1103/PhysRevD.74.035009}{Phys. Rev. D {\normalfont
  \bfseries 74} (2006)  035009},
  \href{http://arxiv.org/abs/hep-ph/0606172}{{\normalfont \ttfamily
  arXiv:hep-ph/0606172}}.

\bibitem{Babu:2015psa}
K.~S. Babu, B.~Bajc, and V.~Susi\v{c}, ``{\em {A minimal supersymmetric E$_{6}$
  unified theory}},'' \href{http://dx.doi.org/10.1007/JHEP05(2015)108}{JHEP
  {\normalfont \bfseries 05} (2015)  108},
  \href{http://arxiv.org/abs/1504.00904}{{\normalfont \ttfamily
  arXiv:1504.00904}}.

\bibitem{Hayreter:2017wra}
A.~Hayreter and G.~Valencia, ``{\em {LHC constraints on color octet
  scalars}},'' \href{http://dx.doi.org/10.1103/PhysRevD.96.035004}{Phys. Rev. D
  {\normalfont \bfseries 96} (2017) no.~3, 035004},
  \href{http://arxiv.org/abs/1703.04164}{{\normalfont \ttfamily
  arXiv:1703.04164}}.

\bibitem{Cheng:2018mkc}
L.~Cheng, O.~Eberhardt, and C.~W. Murphy, ``{\em {Novel theoretical constraints
  for color-octet scalar models}},''
  \href{http://dx.doi.org/10.1088/1674-1137/43/9/093101}{Chin. Phys. C
  {\normalfont \bfseries 43} (2019) no.~9, 093101},
  \href{http://arxiv.org/abs/1808.05824}{{\normalfont \ttfamily
  arXiv:1808.05824}}.

\bibitem{Abel:2008tx}
S.~Abel and V.~V. Khoze, ``{\em {Direct Mediation, Duality and Unification}},''
  \href{http://dx.doi.org/10.1088/1126-6708/2008/11/024}{JHEP {\normalfont
  \bfseries 11} (2008)  024},
  \href{http://arxiv.org/abs/0809.5262}{{\normalfont \ttfamily
  arXiv:0809.5262}}.

\bibitem{Abel:2009bj}
S.~Abel and V.~V. Khoze, ``{\em {Dual unified SU(5)}},''
  \href{http://dx.doi.org/10.1007/JHEP01(2010)006}{JHEP {\normalfont \bfseries
  01} (2010)  006}, \href{http://arxiv.org/abs/0909.4105}{{\normalfont
  \ttfamily arXiv:0909.4105}}.

\bibitem{Kibble:1982ae}
T.~W.~B. Kibble, G.~Lazarides, and Q.~Shafi, ``{\em {Strings in $SO(10)$}},''
  \href{http://dx.doi.org/10.1016/0370-2693(82)90829-2}{Phys. Lett. B
  {\normalfont \bfseries 113} (1982)  237--239}.

\bibitem{Lazarides:2019xai}
G.~Lazarides and Q.~Shafi, ``{\em {Monopoles, Strings, and Necklaces in
  $SO(10)$ and $E_6$}},'' \href{http://dx.doi.org/10.1007/JHEP10(2019)193}{JHEP
  {\normalfont \bfseries 10} (2019)  193},
  \href{http://arxiv.org/abs/1904.06880}{{\normalfont \ttfamily
  arXiv:1904.06880}}.

\bibitem{Lazarides:2021tua}
G.~Lazarides and Q.~Shafi, ``{\em {Triply Charged Monopole and Magnetic
  Quarks}},'' \href{http://dx.doi.org/10.1016/j.physletb.2021.136363}{Phys.
  Lett. B {\normalfont \bfseries 818} (2021)  136363},
  \href{http://arxiv.org/abs/2101.01412}{{\normalfont \ttfamily
  arXiv:2101.01412}}.

\end{thebibliography}\endgroup
\end{document}